\documentstyle[amssymb,twocolumn,prl,aps,graphicx]{revtex}
\begin{document}
\twocolumn[ \hsize\textwidth\columnwidth\hsize\csname
@twocolumnfalse\endcsname

\title{
       Inelastic X-ray scattering study of the collective dynamics in liquid sodium.
      }
\author{
        T.~Scopigno$^{1}$,
        U.~Balucani$^{2}$,
        G.~Ruocco$^{1}$,
        F.~Sette$^{3}$
        }
\address{
    $^{1}$Dipartimento di Fisica and INFM, Universit\'a di Roma ``La Sapienza'', I-00185, Roma, Italy.\\
    $^{2}$Istituto di Elettronica Quantistica CNR, I-50127, Firenze, Italy. \\
    $^{3}$European Synchrotron Radiation Facility, B.P. 220 F-38043 Grenoble, Cedex France.
    }
\date{\today}
\maketitle

\begin{abstract}
Inelastic X-ray scattering data have been collected for liquid
sodium at $T=390$ K, i.e. slightly above the melting point. Owing
to the very high instrumental resolution, pushed up to 1.5 meV,
it has been possible to determine accurately the dynamic
structure factor, $S(Q,\omega)$, in a wide wavevector range, $1.5
\div 15$ nm$^{-1}$, and to investigate on the dynamical processes
underlying the collective dynamics. A detailed analysis of the
lineshape of $S(Q,\omega)$, similarly to other liquid metals,
reveals the co-existence of two different relaxation processes
with slow and fast characteristic timescales respectively. The
present data lead to the conclusion that: {\it i)} the picture of
the relaxation mechanism based on a simple viscoelastic model
fails; {\it ii)} although the comparison with other liquid metals
reveals similar behavior, the data do not exhibit an exact scaling
law as the principle of corresponding state would predict.
\end{abstract}


]

\section{INTRODUCTION}

Liquid metals are well known to exhibit remarkably pronounced
inelastic features (Brillouin peaks) in their density fluctuations
spectra up to very high $Q$ values (well beyond the hydrodynamic
regime). For this main reason, in the last decades, the dynamic
structure factor of these systems has been used as a benchmark for
the comprehension of the mechanisms underlying the atomic motions
at the microscopic level in liquids. In particular, in the special
case of alkali metals, collective oscillatory modes persist down
to wavelengths of the order of one or two interparticle distances,
a feature which makes them ideal candidates to test different
models for collective properties at finite wavevectors.

In the last thirty years many experimental investigations have
been performed on these systems by means of inelastic neutron
scattering (INS), which was, up to a few years ago, the only tool
adequate to investigate the dynamic of condensed matter in the
mesoscopic wavevector region. Indeed, since the very first
inelastic neutron scattering experiments by Copley and Rowe
\cite{rubidio2} in liquid rubidium, many similar studies have been
reported for similar systems: lithium \cite{verkerk}, sodium
\cite{sodio}, potassium \cite{potassio}, cesium \cite{cesio} and
again rubidium \cite{rubidio}. In the most favorable cases (for
example in cesium), important features such as the so called {\it
positive dispersion} of sound, namely an upwards deviation from
the low-$Q$, linear dispersion behavior, have been reported.
Moreover, a description of the coherent dynamic structure factor
has been given in terms of a single relaxation process
(viscoelastic model). On the other hand, in some of the above
mentioned systems, an accurate investigation of the collective
properties is prevented by two drawbacks of INS technique, namely
the presence of both coherent and incoherent contribution to the
inelastic scattering cross section, and the occurrence of
kinematic restrictions which confine the minimum accessible
wavevector above a threshold value which is already outside the
linear dispersion regime. This is particularly the case of lighter
elements such as lithium and sodium.

Despite the lack of experimental information on the collective
properties in these systems, significant advances have been done
by means of computer simulation techniques, where the afore
mentioned restrictions do not apply. Consequently, many numerical
studies have been reported aiming at the study of the density
fluctuation spectra of liquid metals
\cite{rahman1,umblitiomd,MDworks,kamb,canaleslitiomd,dm1,dm2}.
Moreover, by these numerical techniques one can access a wider set
of correlation functions, while an inelastic scattering
experiment basically probes the density autocorrelation function.

On the theoretical side, thanks to development of approaches such
as the memory function formalism, the relaxation concept, the
kinetic theory,
\cite{libroumberto,BY,mori,leves,GOT,sjogr,desh,beng,casasMC,patacca,yulmetyev}
it has been built up a framework able to account for the behavior
of the afore mentioned correlation functions. Along this line, a
general picture of the dynamics of the density fluctuations in
simple liquids gradually emerged. In particular it is now
established that the decay of the density autocorrelation function
occurs through mechanisms characterized by different time scales.
In addition to a first mechanism related to the coupling of the
density and the temperature modes (thermal relaxation), one must
allow for the dynamical effects of the stress correlation function
(viscous relaxation). This latter process is thought to proceed
throughout two distinct relaxation channels, active over quite
different time scales: a first rapid decay, customarily ascribed
to the interaction of each atom and the "cage" of its neighboring
atoms, is followed by a slower process which yields a long lasting
tail. This mechanism stems from slow, temperature-dependent
rearrangements which eventually, in those systems capable to
sustain supercooling, may cause a structural arrest (glass
transition). This structural relaxation process has widely been
studied within the mode coupling formalism. Recently, in the
specific case of liquid alkali metals, remarkable attempts have
been made to set up self consistent approaches which include all
these mechanisms from the very start \cite{casasMC,casasMC1}.

This theoretical framework has been tested on a number of
numerical studies (see, for example, \cite{libroumberto}). On the
experimental side, however, the lineshapes extracted by INS in
those systems where one may access a significant ($Q-E$) region
did not allow to discriminate between different models including
all the relevant relaxation processes \cite{cesio}.

Although trough of INS enormous advances in the comprehension of
the collective properties of condensed matter have been reported,
some basic aspects remain still unsettled. In particular, due to
the previously mentioned features of INS, sufficiently accurate
measurements of the coherent dynamic structure factor,
$S(Q,\omega)$ have hardly been reported, with the result that a
quantitative description of the relaxation dynamics in monoatomic
fluids was mainly based on simulation data.

Only in the recent past, the development of new synchrotron
radiation facilities opened the possibility of using X-rays to
measure $S(Q,\omega)$ in the non-hydrodynamic region; in this
case the photon speed is obviously much larger than the velocity
of the excitations, and no kinematic restriction occurs. Moreover,
in a monatomic system, the X-ray scattering cross section is
purely coherent, and consequently can be directly associated with
the coherent dynamic structure factor \cite{libroumberto,loves}.
After some early, pioneering data \cite{burk}, IXS rapidly
developed in the last decade, becoming fully operational with the
advent of the third generation sources \cite{noiV,noiM}. The
applicability of IXS to the study of collective and single
particle dynamics of light liquid metals has been then exploited
in several works performed at the beamline ID16 of the european
synchrotron radiation facility (ESRF) in Grenoble
\cite{sinn,euro,pilgrim}, thus showing the complementarity of IXS
and INS. More recently, IXS has been pushed to the top of its
theoretical performance in terms of flux and resolution, and
therefore new and more accurate IXS studies on liquid metals
allowed to explore the relaxation processes underlying the
microscopic dynamics in monoatomic fluids \cite{jpc,PRL,PRE}. In
particular, it has been possible to detect experimentally and to
quantify the presence of the relaxation scenario predicted by the
theory of simple liquids. In detail, the data have been analyzed
within a generalized hydrodynamic approach \cite{libroumberto,BY}
and, following a memory function formalism \cite{mori} in the
mesoscopic wavevector region, it has been possible to detect the
presence of two viscous relaxation processes in addition to the
thermal process \cite{PRL}. These results, obtained with a
phenomenological ansatz for the memory function originally
proposed in a theoretical work by Levesque et al. \cite{leves},
has then been deeply discussed in order to clarify the physical
origin of all these dynamical processes \cite{jpc}.

In this work, we report the results of a very high resolution
determination of the dynamic structure factor in liquid sodium at
the melting point at several fixed $Q$ values. Several
approximation are adopted for the analysis of the experimental
data, and the number and the role of the involved relaxation
processes is discussed in detail. Finally, a comparison with
similar experimental data on different liquid metals is presented.

\section{The Experiment}

In liquid sodium, the coherent and incoherent INS cross sections
are almost equivalent, and the adiabatic speed of sound is about
2500 m/s. As a result, a study of the collective properties in the
Q-region characterized by a linear dispersion (the first
Brillouin pseudo-zone) is quite hard to perform \cite{sodio}. The
use of IXS to these systems is, therefore, particularly
advantageous and, as shown in ref. \cite{pilgrim}, the inelastic
features of the dynamic structure factor can easily be detected
even around the Q position of the first sharp diffraction peak
($Q_m \approx 20$ nm$^{-1}$).

Data have been collected for liquid sodium at the new IXS beamline
ID28 of ESRF (which is now fully operative), allowing an energy
resolution of $\Delta E=1.5$ meV FWHM. The experiment has been
performed at fixed exchanged wavevector over a $Q-$ region $1.5
\div 15$ nm$^{-1}$. Each energy scan ($-50 < E < 50$ meV) took
about 300 minutes, and has been repeated for a total integration
time of about 500 seconds/point. A five analyzers bench,
operating in horizontal scattering geometry, allowed us to
collect simultaneously photons at five different values of
exchanged wavevector $Q$ for each single scan. As sample
environment we used a setup similar to the one adopted in previous
experiments performed on liquid lithium \cite{euro}, namely an
austenitic steel cell heated by thermal contact with a resistor
connected to a voltage regulated supply. The sample length has
been optimized by matching the absorption length of sodium at the
working energy 21747 eV (corresponding to the Si (11 11 11)
reflection in backscattering geometry) which is about 5 mm.

\begin{figure}[p]
\centering
\includegraphics[width=.55\textwidth]{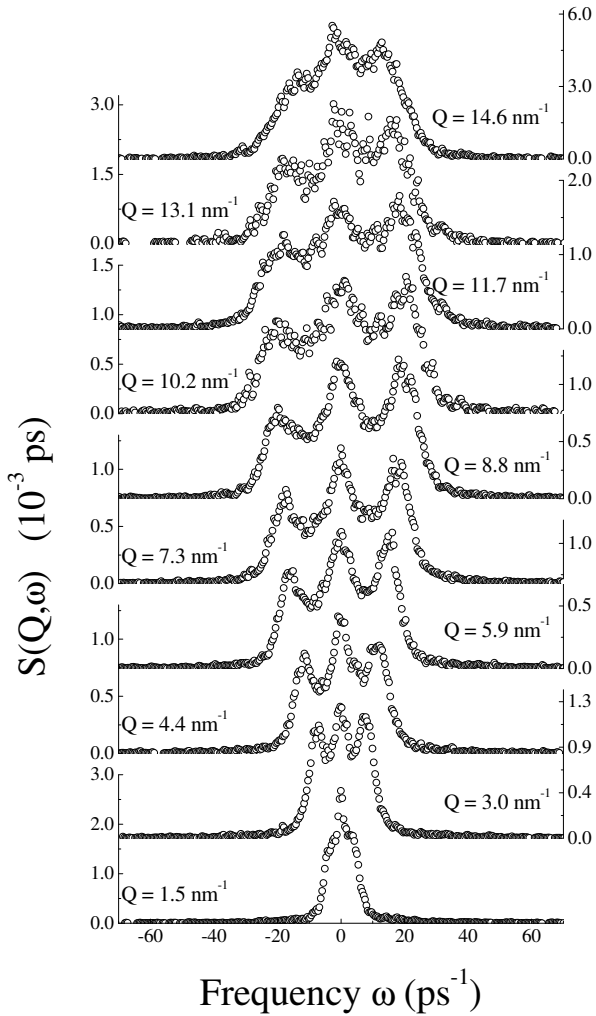}
\caption{Selected IXS energy spectra of liquid sodium at $T=390$ K
and fixed values of exchanged wavevector (open dots $\circ$). The
data have been normalized utilizing the sum rules according to the
procedure described in the text.} \label{napanel}
\end{figure}

In Fig. \ref{napanel}, a selection of experimental spectra is
reported. Since the incident flux on the sample varies with time,
data have been normalized to the monitor. Moreover, since the five
analyzers have different scattering efficiencies, to extract
$\widetilde{S}(Q,\omega)$ from the raw spectra the data have been
put in absolute units using the knowledge of the lower order
frequency momenta of the dynamic structure factor:

\begin{eqnarray}
\Omega_{\widetilde{S}}^{(0)}=\int d\omega
\widetilde{S}(Q,\omega)=\widetilde{S}(Q) \\
\Omega_{\widetilde{S}}^{(1)}=\int d\omega
\widetilde{S}(Q,\omega)\omega=\frac{\hbar Q^{2}}{2M}
\end{eqnarray}

\noindent where the tilde indicates the true quantum dynamic
structure factor, to distinguish it from its classical
representation.

This procedure can be summarized as follows. The actual
experimental spectrum $I(Q,\omega)$ is proportional to the
convolution of the scattering law $\widetilde{S}(Q,\omega)$ with
the resolution $R(\omega)$:

\begin{equation}
I(Q,\omega )=E(Q)\int d\omega ^{\prime }\widetilde{S}(Q,\omega
^{\prime })R(\omega -\omega ^{\prime })  \label{convo}
\end{equation}

\noindent In terms of the first two spectral moments of $I$ and
$R$: $\Omega_{I}^{(0)}$, $\Omega _{I}^{(1)}$, $\Omega _{R}^{(0)}$,
$\Omega _{R}^{(1)}$, one easily deduces that

\begin{equation}
\widetilde{S}(Q)=\frac{\hbar Q^{2}}{2M}(\Omega _{I}^{(1)}/\Omega
_{I}^{(0)}-\Omega _{R}^{(1)}/\Omega _{R}^{(0)})^{-1}.
\label{norma}
\end{equation}

\noindent Consequently, the normalized spectrum (still affected by
the resolution broadening) reads:

\begin{equation}
I_N(Q,\omega )=\frac{\widetilde{S}(Q)}{\int I(Q,\omega d\omega)}
I(Q,\omega )
\end{equation}

\noindent The reliability of this procedure will be discussed in
the next section.

\section{Data Analysis}

The data analysis has been performed following a memory function
approach, similar to the one adopted for lithium and aluminum
data \cite{jpc,PRL,PRE}.

Within the generalized Langevin equation formalism, it is possible
to express the \textit{classical} dynamic structure factor in
terms of a complex function, $M(Q,t)$, (memory function) related
to the interaction details. In particular, $S(Q,\omega)$ can be
expressed in terms of the real and imaginary part of
$\tilde{M}(Q,\omega)$, the Fourier-Laplace transform of $M(Q,t)$
as \cite{libroumberto}:

\begin{equation}
\frac{S(Q,\omega )}{S(Q)}= \frac{\pi^{-1}\omega _0^2(Q)\tilde{M}^{\prime }(Q,\omega )}{%
\left[ \omega ^2-\omega _0^2(Q)+\omega \tilde{M}^{\prime \prime
}(Q,\omega )\right] ^2+\left[ \omega \tilde{M}^{\prime }(Q,\omega
)\right] ^2} \label{sqwgenerale}
\end{equation}

\noindent where the quantity $\omega _0^2(Q)=\frac{KTQ^2}{mS(Q)}$
is related to the generalized isothermal sound speed through the
relation $c_t(Q)=\omega_0(Q)/Q$.

\noindent In order to use the above expression to reproduce the
experimental spectra, $I_N(Q,\omega )$, it has to be modified to
be quantum-compliant with the detailed balance condition, and
finally must be convoluted with the instrumental resolution
$R(\omega)$. Utilizing one of the most common quantum
transformation, one finds that

\begin{equation}
I^{th}_N(Q,\omega)=\int\frac{\hbar \omega' / KT }{1-e^{-\hbar
\omega' / KT }}S(Q,\omega' )R(\omega- \omega')d\omega '
\label{fitfunction}
\end{equation}

The above expression can finally be used as a fitting function to
the experimental data to obtain the relevant relaxation
parameters, i.e. relaxation times and strengths. In this
procedure, the values of $\widetilde{S}(Q)$ are extracted from the
experimental data exploiting Eq. (\ref{norma}).

\noindent As memory function, $M(Q,t)$, we utilized a multiple
exponential ansatz. In Ref. \cite{PRL}, indeed, it has been shown
that in a similar system, namely liquid lithium: {\it i)} at least
three distinct relaxation processes (one thermal and two viscous)
are necessary to reproduce the IXS lineshape; and {\it ii)} a
multi-exponential shape is an acceptable approximation of the
actual memory function. Consequently, the total memory function
reads

\begin{eqnarray}
M(Q,t) &=&\left( \gamma -1\right) \omega _{0}^{2}(Q)e^{-D_{T}
Q^{2}t}  \label{mem} \\ &+&\Delta ^{2}(Q)\left[ A(Q)e^{-t/\tau
_{\alpha }(Q)}+(1-A(Q))e^{-t/\tau _{\mu }(Q)}\right] \nonumber
\end{eqnarray}

\noindent The first term of the above equation comes from the
coupling between thermal and density degrees of freedom, and
depends on the thermal conductivity $D_T$ and the specific heat
ratio $\gamma$. The remaining two contributions in Eq. (\ref{mem})
are the genuine viscous processes. Through the fitting procedure,
the unknown parameters $\tau_\mu, \tau_\alpha$ (the relaxation
times), $A(Q)$ (the relative weight of the viscous processes),
and $\Delta ^{2}(Q)$ (the total viscous strength), can be deduced.

\begin{figure}[h]
\centering
\includegraphics[width=.52\textwidth]{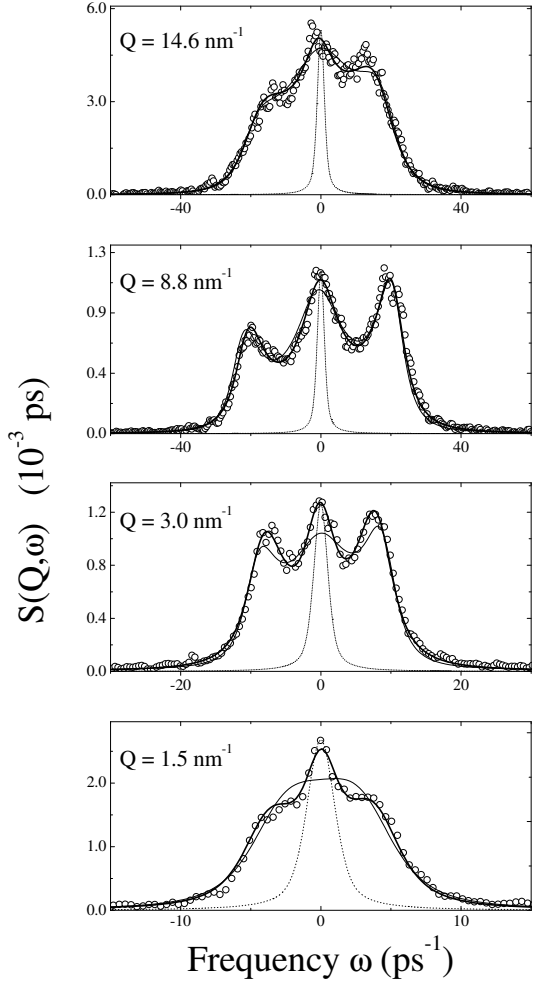}
\vspace{-.1cm} \caption{IXS spectra (open dots $\circ$) fitted
with one (thin line) or two (thick line) relaxation process.
Experimental resolution is also indicated (dotted line)}
\label{fit}
\end{figure}

In Fig. \ref{fit}, we report the outcome of this fitting
procedure, i.e. a comparison between $I^{th}_N(Q,\omega)$ and
$I_N(Q,\omega)$. Both the results of a one and a two time ansatz
are shown. As in the case of lithium and aluminum, a single
relaxation time description (viscoelastic model) appears to yield
a rather poor reproduction of the data. In particular, at low
$Q$'s the simultaneous presence of two processes is crucial even
to catch just the qualitative features of the spectra.

As previously mentioned, once that the experimental values of
$\widetilde{S}(Q)$ are introduced in Eq. (\ref{sqwgenerale}), the
only free fitting parameters are the relaxation times and
strength. A consistency check has however been obtained comparing
these values of $\widetilde{S}(Q)$ (sum rules) with the one
calculated from $\omega^2_0(Q)$ left as free parameter in Eq.
(\ref{sqwgenerale}).

\begin{figure}[h]
\centering
\includegraphics[width=.57\textwidth]{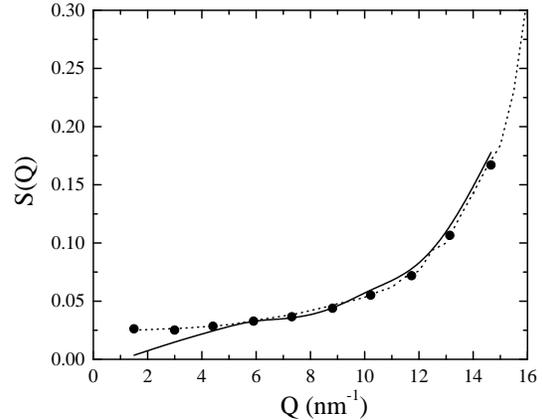}
\vspace{-1.3cm} \caption{Values of S(Q) from the spectral moments
of the IXS signal (full line), compared with the values taken from
the fit (full circles $\circ$) and from literature (dotted line
$\cdot\cdot\cdot$)} \label{normal}
\end{figure}

\begin{figure}[h]
\centering \vspace{-.8cm}
\includegraphics[width=.4\textwidth]{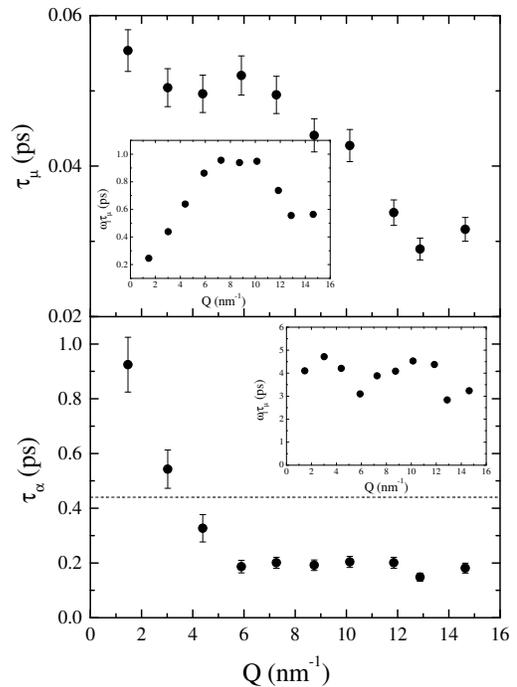}
\vspace{-.1cm} \caption{Main frames: Fast and slow relaxation
times as obtained by the fitting procedure. Insets: product of the
relaxation times with the position of the inelastic peak of the
classical density correlation function, as deduced by the fit.}
\label{times}
\end{figure}

In Fig. \ref{normal} the results are shown, together with the
experimental determination of $S(Q)$ from neutron diffraction
data \cite{ose}.

In Fig. \ref{times} we show the two relaxation times,
$\tau_\alpha$ and $\tau_\mu$ as obtained by the fitting procedure.
The slower process is almost constant, with an abrupt increase
below $Q \approx 6 $ nm$^{-1}$. Such an increase is probably an
artifact, since it occurs as the timescale approaches the inverse
of the resolution width, here reported as a dashed line.
Consequently the determination of the $\alpha$ relaxation time
below this $Q$ value is no longer reliable. The $\mu$ relaxation
process shows instead a slightly decreasing $Q$ dependence. In the
insets, we report $\omega \tau$ as function of $Q$ for the two
processes, where omega is the dominant frequency of the density
fluctuations at a given $Q$ (i.e. the frequency of the Brillouin
component). The quantity $\omega \tau$ gives important indications
about the role of the two processes. The faster one, indeed, has
values of $\omega \tau$ always smaller than one, so that it is
mainly responsible for the acoustic damping (Brillouin linewidth).
Moreover, $\omega(Q) \tau_\mu(Q)$ shows an increase in the same
wavevector region where the sound dispersion occurs
(approximately, up to $Q$ values of one half of the position of
the first sharp diffraction peak), so that an increase of the
apparent peak position ruled by the strength of the fast process
has to be expected. The $\omega(Q) \tau_\alpha(Q)$ is always
larger than one. Consequently the alpha process mainly control
the sharper portion of the quasielastic scattering, while it only
gives a moderate contribution to the speed of sound over all the
explored $Q$ range (the strength of the $\alpha$ process is
negligible with respect to that of the fast relaxation).

\begin{figure}[h]
\centering
\includegraphics[width=.45\textwidth]{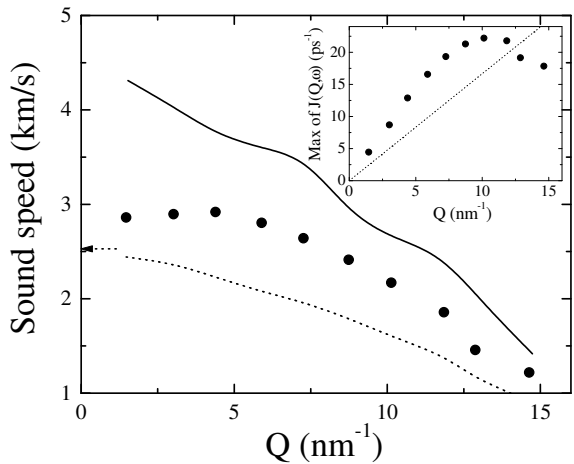}
\vspace{-4.7cm} \caption{Inset: dispersion curve $\omega_l (Q)$
from the deconvoluted lineshapes (full dots $\bullet$) compared
with the low frequency, isothermal, behavior (dotted line $\cdot
\cdot \cdot$). Main plot: speed of sound in the same notation of
the inset. The instantaneous value $c_{\infty} (Q)$ (full line) is
also reported. The $c_0(Q\rightarrow 0)$ limit of the isothermal
value, from compressibility data, is shown as dotted arrow.}
\label{dispersion}
\end{figure}

The above mentioned increase of the speed of sound is illustrated
in Fig. \ref{dispersion}, where $\omega_l (Q)$, the maximum of the
longitudinal current correlation spectrum
$J^{L}(Q,\omega)=\omega^2/Q^2S(Q,\omega)$ is reported in the main
frame, and the associated sound speed $\omega_l (Q)/Q$ is reported
in the inset. The positive dispersion turns out to be of the order
of $20 \%$, similarly to what is found in other liquid metals
\cite{libroumberto}. In contrast with the results reported in
Ref.\cite{pilgrim}, we find that the isothermal value of the sound
speed is not yet reached at the minimum wavevector probed in our
experiment, i.e. 1.5 nm$^{-1}$. At this low $Q$ resolution
effects are particularly important: even with the present
resolution (1.5 meV FWHM) the Brillouin components of the
$S(Q,\omega)$ appear as shoulders of the quasielastic signal,
therefore we believe that this discrepancy may be ascribed to the
lower resolutions of the previous experiment that prevents an
accurate determination of the actual lineshape at such small $Q$
value. Both the isothermal $c_t(Q)$ (low frequency) \cite{notact}
and the infinite frequency $c_\infty (Q)$ values of the sound
speed have been reported as obtained by the fit. This latter
quantity is always higher than the maximum value reached by the
apparent sound speed, consistently with the behavior of the
relaxation times: as shown in Fig.\ref{times}, the faster process
(which is the one with the dominant strength) has a relaxation
time which never exceeds the dominant timescale of the density
fluctuation at the probed wavevector. This notwithstanding, the
actual values of $c_\infty (Q)$ reported in Fig. \ref{dispersion}
can partially be affected by the oversimplifications given by the
assumption of exponential lineshape for the memory functions
\cite{jpc}.

\begin{figure}[h]
\centering \vspace{-.9cm}
\includegraphics[width=.5\textwidth]{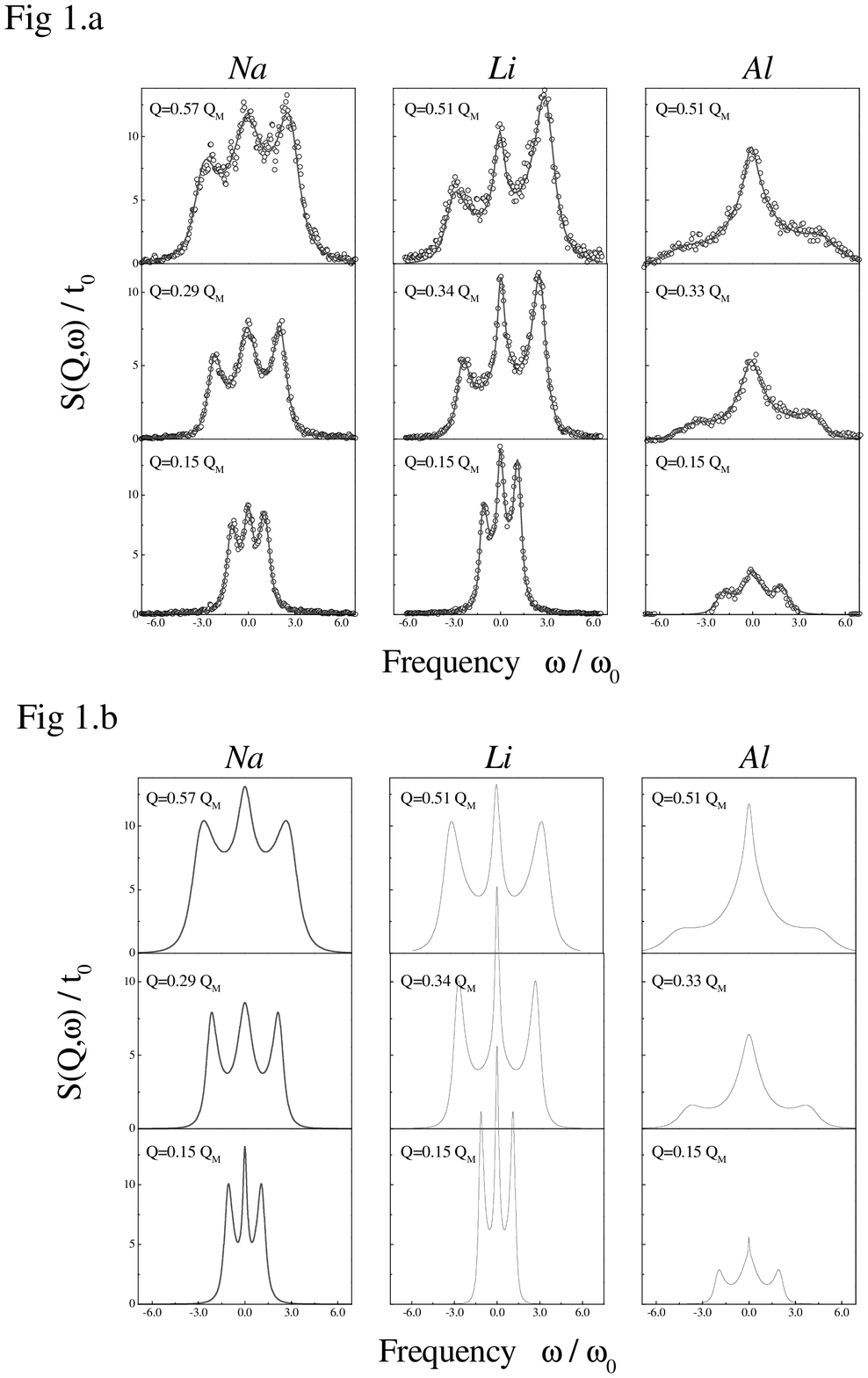}
\vspace{-.5cm} \caption{Comparison of the IXS spectra of sodium,
lithium and aluminium reported in scaled units, as defined in the
text. Best fitted lineshape (two time relaxation model) are also
shown. b) same as a) but with a classical, resolution
deconvoluted, lineshape as obtained from the values of the
parameters deduced by the fitting procedure.} \label{comparison}
\end{figure}

We conclude this investigation with the comparison of the
$S(Q,\omega)$ of three different liquid metals. In Fig.
\ref{comparison} the IXS data on liquid sodium are compared with
those of lithium and aluminium by introducing a set of length,
mass and time units $[l^*,m^*,t^*]$ for each system. Such a
procedure can be considered meaningful as far as is valid a
corresponding state principle. As length unit, $l^*$, we have
chosen $l^*=Q_m^{-1}$, the inverse of the position of the static
structure factor S(Q). The time unit, $t^*$, has been instead
chosen as $t^*=\sqrt{\frac{m}{T_m}}l^*$, with $T_m$ the melting
temperature and $m$ the atomic mass. In the upper panel the raw
IXS signal, scaled on this units is reported. The $Q$ values have
been selected among the available ones as closer as possible for
three different regions of the dispersion curve. It is worth to
point out that all the spectra are resolution broadened and,
moreover, that the three experiments have been performed in
different condition of resolutions. To bypass the instrumental
resolution bias, one can compare the lineshape obtained by the
fits, i.e. the classical deconvoluted dynamical structure factor
that better reproduces the experimental data after the
transformation of Eq. (\ref{fitfunction}). The result of such a
comparison is reported in the lower panel of Fig.
\ref{comparison}. The aluminum spectra clearly have a different
lineshape, while lithium and sodium are much more similar when
scaled in their respective absolute units. This notwithstanding,
the lineshapes of the two alkali systems do not coincide,
implying that the principle of corresponding state is not
entirely valid for these systems. One comes to the same conclusion
even looking at some basic quantities such as the sound speed:
the scaling factors between lithium and sodium, for example, are
not the same if one looks at $c_t(Q\rightarrow 0)$ or
$c_\infty(Q\rightarrow 0)$. Also the static structure factor
$S(Q)$, rather of being universal quantity, depends on the
considered system.

\section{CONCLUSIONS}

In conclusion, we presented a very high resolution experimental
study of the collective dynamic structure factor in liquid sodium
at the melting temperature. In order to account for all the
spectral details, beside the usual thermal process coupling
temperature and density, at least two additional viscous
relaxation processes have to be invoked. The viscoelastic
approximation do not give, indeed, a satisfactory description of
our data. These results have finally been compared with those
obtained in other liquid metals. Although the mechanisms ruling
the dynamics turn out to lie in a general framework
characterizing both alkali and non-alkali simple liquids, some
important quantitative differences are found: in particular, a
scaling law among alkali systems seems to be unapplicable, in
contrast to some previous findings obtained by computer
simulation data.

\section{ACKNOWLEDGEMENTS}

We are thankful to the ESRF staff for the great experimental
conditions and for the assistance during the experiment. One of
the authors (T.S.) acknowledges R. Yulmetyev and M. Silbert for
stimulating discussions.


\begin{references}
\bibitem {notact}   Due to the high thermal conductivity of liquid metals as
compared to that of ordinary fluids, the crossover in the sound
speed associated with the thermal relaxation occurs well below 1
nm$^{-1}$, which is approximately the lower $Q$ accessible to
IXS. Consequently, in the IXS window the low frequency limit of
the sound speed is the isothermal rather than the adiabatic sound
speed. \cite{jpc}
\end{references}
\end{document}